\pdfoutput=1
\documentclass[superscriptaddress,twocolumn,aps,preprintnumbers,amsmath,amssymb,nofootinbib]{revtex4-1}

\usepackage{epsfig}
\usepackage{amsmath}
\usepackage{amssymb}
\usepackage{subfigure}
\usepackage{cancel}
\usepackage{textcomp}
\usepackage{calc}
\usepackage{graphicx}
\usepackage{dcolumn}
\usepackage{color}
\usepackage{xcolor}
\usepackage{pifont}
\usepackage{slashed}
\usepackage{bbold}
\usepackage{xfrac}
\usepackage{orcidlink}
\usepackage{comment}
\usepackage{enumitem}

\usepackage{hyperref}
\hypersetup{colorlinks=true,urlcolor=blue,linkcolor=red,citecolor=green!60!black}

\begin{document}

%%%%%%%%%%%%%%%%%%%%%%%%%%%%%%%%%%%%%%%%%%%%%%%%%%%%%%%%%%%%%%%%%%%%%%%%%%%%%%%%%%%%%%%%%%%%%%%%%%%%

\preprint{MS-TP-24-25}
\preprint{MITP-24-074}

\title{Gravitational charge production}

\author{Martin A. Mojahed \orcidlink{0000-0002-5591-7364}}
\email{mojahedm@uni-mainz.de}
\affiliation{Physik-Department T70, Technische Universit\"{a}t M\"{u}nchen,\\
James-Franck-Straße 1, D-85748 Garching, Germany}
\affiliation{PRISMA$^+$ Cluster of Excellence and Mainz Institute for Theoretical Physics\\
Johannes Gutenberg University, 55099 Mainz, Germany}

\author{Kai Schmitz \orcidlink{0000-0003-2807-6472}}
\email{kai.schmitz@uni-muenster.de}
\affiliation{Institute for Theoretical Physics, University of M\"unster, 48149 M\"unster, Germany}

\author{Xun-Jie Xu \orcidlink{0000-0003-3181-1386}}
\email{xuxj@ihep.ac.cn}
\affiliation{Institute of High Energy Physics, Chinese Academy of Sciences, Beijing 100049, China}

%%%%%%%%%%%%%%%%%%%%%%%%%%%%%%%%%%%%%%%%%%%%%%%%%%%%%%%%%%%%%%%%%%%%%%%%%%%%%%%%%%%%%%%%%%%%%%%%%%%%

\begin{abstract}
Wash-in leptogenesis is an attractive mechanism to produce the baryon asymmetry of the Universe. It treats right-handed-neutrino interactions as spectator processes, on the same footing as electroweak sphalerons, that reprocess primordial charge asymmetries in the thermal plasma into a baryon-minus-lepton asymmetry. The origin of these primordial charges must be accounted for by new $CP$-violating dynamics at very high energies. In this paper, we propose such a scenario of chargegenesis that, unlike earlier proposals, primarily relies on new interactions in the gravitational sector. We point out that a coupling of a conserved current to the divergence of the Ricci scalar during reheating can lead to nonzero effective chemical potentials in the plasma that, together with a suitable charge-violating interaction, can result in the production of a primordial charge asymmetry. Gravitational chargegenesis represents a substantial generalization of the idea of gravitational baryogenesis. We provide a detailed analysis of a generic and minimal realization that is consistent with inflation and show that it can successfully explain the baryon asymmetry of the Universe.

\end{abstract}

%%%%%%%%%%%%%%%%%%%%%%%%%%%%%%%%%%%%%%%%%%%%%%%%%%%%%%%%%%%%%%%%%%%%%%%%%%%%%%%%%%%%%%%%%%%%%%%%%%%%

\date{\today}
\maketitle

%%%%%%%%%%%%%%%%%%%%%%%%%%%%%%%%%%%%%%%%%%%%%%%%%%%%%%%%%%%%%%%%%%%%%%%%%%%%%%%%%%%%%%%%%%%%%%%%%%%%

\section{Introduction}

The excess of matter over antimatter in our Universe~\cite{Planck:2018vyg} provides evidence for new physics beyond the Standard Model (SM), and its origin remains an open question. A popular class of mechanisms for dynamically generating the baryon asymmetry in the early Universe (BAU) is known as leptogenesis~\cite{Fukugita:1986hr}, which requires an extension of 
the SM field content by right-handed neutrinos (RHNs).
Most realizations of the leptogenesis idea rely on both charge--parity ($CP$) violation and out-of-equilibrium $B\!-\!L$-violating interactions\,---\,two essential ingredients for the generation of the BAU \,---\,to emerge from the RHN sector~\cite{Bodeker:2020ghk}. 
The recent wash-in leptogenesis proposal~\cite{Domcke:2020quw,Domcke:2022kfs}, however, is an exception to this lore, which allows a hierarchy between the temperature scales of $CP$ and $B\!-\!L$ violation and does not require any $CP$ violation from the RHN sector. 

At the core of wash-in leptogenesis lies the idea of considering non-trivial chemical background configurations in the early Universe and treating RHNs on par with electroweak (EW) sphalerons. Analogous to how sphalerons act on the chemical background to wash in a $B\!+\!L$ asymmetry, the RHNs act as spectator fields providing a new equilibrium attractor for the chemical potentials that generically features nonzero $B\!-\!L$, even if $B\!-\!L=0$ initially~\cite{Domcke:2020quw}. As the wash-in leptogenesis mechanism requires a non-trivial chemical background as an initial condition to wash in a $B\!-\!L$ asymmetry, it only qualifies as a full theory of baryogenesis when accompanied by ultraviolet (UV) dynamics that can generate the chemical background. The UV completion of wash-in leptogenesis may be referred to as {\it chargegenesis}: a mechanism that provides the necessary $CP$-violating initial conditions for wash-in leptogenesis, but which itself may still conserve the $B\!-\!L$ charge of the Universe. 

Thus far, several scenarios of chargegenesis have been studied in the literature, including axion inflation~\cite{Domcke:2022kfs}, heavy Higgs decays~\cite{Mukaida:2024eqi}, and the evaporation of primordial black holes~\cite{Schmitz:2023pfy}. Common to all of these studies is that they rely on particle dynamics. In the present work, we shall propose a new avenue for chargegenesis\,---\,{\it gravitational chargegenesis}\,---\,in which gravity takes center stage.

Our gravitational chargegenesis setup shares similarities with gravitational baryogenesis, which is based on an effective interaction of the following form~\cite{Davoudiasl:2004gf}
\begin{align}
    \label{riccicurrentGeneral}
    {\cal L}\supset \frac{1}{M_*^2}\int d^4x\,\sqrt{-g}\,J_A^\mu\,\partial_\mu \mathcal{R},
\end{align}
where $\mathcal{R}$ is the Ricci scalar,  $M_*$ parametrizes the UV scale relevant to the generation of this interaction, and $J_A$ denotes a generic current, which in gravitational baryogenesis is the baryon current. In our mechanism, by contrast, $A$ can be any of the SM global charges that are conserved in the SM plasma at high temperatures and that can be employed in wash-in leptogenesis.

In the Friedmann–Lema\^itre–Robertson–Walker (FLRW) Universe, 
Eq.~(\ref{riccicurrentGeneral}) leads
to an effective chemical potential associated with the current $J_A$,
\begin{equation}
	\mu_{A}^{{\rm eff}} = \frac{\dot{{\cal R}}}{M_{*}^{2}}=\frac{3\rho\dot{\omega}}{M_{*}^{2} M_{P}^{2}}+\frac{\sqrt{3}(1+\omega)(1-3\omega)}{M_{*}^{2}M_{P}^{3}}\rho^{\frac{3}{2}}\thinspace,\label{eq:chem}
\end{equation}
where $\dot{\mathcal{R}}$ is the time derivative of the Ricci scalar, $M_P$ is the reduced Planck mass, $\rho$ denotes the energy density of the Universe, and $\omega$
is the equation-of-state (EOS) parameter, i.e.\ the ratio of pressure
and energy density. The effective chemical potential $\mu_{A}^{{\rm eff}}$ induces non-vanishing chemical potentials $\mu_i$ for all particle species $i$ carrying an $A$ charge. In general, these particle species also carry other global charges. If one of these charges, say $C$, should be violated by an interaction in the plasma, the bias introduced by the chemical potentials $\mu_i \propto \mu_{A}^{{\rm eff}}$ will hence lead to the generation of a nonzero charge asymmetry in equilibrium, $q_C^{\rm eq}$. The exact relation between $q_C^{\rm eq}$ and $\mu_{A}^{{\rm eff}}$ depends on the susceptibility matrix of the plasma, $\chi_{CA}$, see Eq.~(2.31) in Ref.~\cite{Domcke:2020axion} for an explicit expression that can be readily applied to our scenario.

While the origin of the interaction in Eq.~\eqref{riccicurrentGeneral} was originally envisioned to arise from quantum gravity, it was later argued that an effective interaction of this form is a generic outcome in models that feature high-scale $CP$ violation in a curved spacetime background, which can result in a cutoff scale $M_* \ll M_P$~\cite{McDonald:2014yfg,McDonald:2015iwt,McDonald:2016ehm,McDonald:2020ghc}. In fact, a theory of gravitational leptogenesis was studied previously, where $A$ was identified with lepton number~\cite{McDonald:2014yfg,McDonald:2015iwt,McDonald:2016ehm,McDonald:2020ghc,Samanta:2020cdk,Samanta:2020tcl,Samanta:2021zzk}. It was shown that the interaction~\eqref{riccicurrentGeneral} is generated dynamically at two loops in the type-I seesaw mechanism~\cite{Minkowski:1977sc,Yanagida:1979as,Gell-Mann:1979vob,Mohapatra:1979ia} when the seesaw Lagrangian is minimally coupled to the gravitational background as a direct consequence of $CP$ violation in the RHN Yukawa sector. When combined with lepton-number violation in the RHN sector, this provided a new contribution to leptogenesis. Here, we point out that the charge $A$ associated with the current in Eq.~\eqref{riccicurrentGeneral} and the charge $C$ that is being violated do \textit{not} have to be the same charge, which relaxes the assumptions required for successful baryogenesis via chargegenesis. Additionally, the RHN sector is liberated from $CP$ violation constraints, as the RHNs are only required to wash in a nonzero $B-L$ asymmetry.

The effective chemical potential in Eq.~\eqref{eq:chem} 
vanishes if $\omega=-1$ or $1/3$, which implies that when the Universe
is dominated by vacuum energy (e.g., during inflation)
or radiation, the gravitationally generated chemical potential is
suppressed. 
The suppression can be alleviated by enhancing the trace anomaly of the energy--momentum tensor, e.g., by adding many new particles charged under a new $SU(N)$ gauge group to the thermal plasma~\cite{Davoudiasl:2004gf}. Alternatively, one may replace
%a non-minimal choice of $f(\mathcal{R})$ 
$\mathcal{R}$ in Eq.~\eqref{riccicurrentGeneral} by a scalar function $f(\mathcal{R})$~\cite{Li:2004hh,Feng:2004mq}, or modified theories of gravity~\cite{Lambiase:2006dq,Lambiase:2013haa,Oikonomou:2016jjh,Odintsov:2016hgc,Ramos:2017cot,Bhattacharjee:2020wbh,Pereira:2024ddu,Pereira:2024kmj}, see also~\cite{Sadjadi:2007dx} for earlier work and Refs.~\cite{DeSimone:2016bok,Liang:2019fkj} for related ideas. 

In the present work, we consider a simple framework to evade the large suppression of $\dot{\mathcal{R}}$ that relies on nothing but the oscillations of the inflaton field $\phi$ during the stage of reheating after inflation. After a period of slow-roll inflation driven by $\phi$ rolling down its potential $V(\phi)$, $\phi$ starts oscillating around the minimum of its potential. If the potential can be approximated as a monomial around its minimum, $V(\phi)\approx \phi^p$, the EOS parameter reads $\omega\approx(p-2)/(p+2)$~\cite{Turner:1983he}, which in general differs from $1/3$ or $-1$. 
As we will show, this renders the gravitational chargegenesis mechanism compatible with high-scale inflationary setups, without modifying gravity or adding extra particle species to the thermal plasma. We also provide a detailed time-resolved picture of how the charge asymmetry is generated as the Universe transitions from a period of inflation to reheating and beyond. Throughout this work, we employ Planck units, $M_{P}=1$.

\section{Equations and solutions}

The evolution of the inflaton field $\phi$, the radiation energy density $\rho_{R}$,
and the charge asymmetry $q_{C}$ is described by the following set of equations,
\begin{align}
	\ddot{\phi}+\left(3H+\Gamma_{\phi}\right)\dot{\phi} & =-\frac{dV}{d\phi}\thinspace,\label{eq:phi}\\
	\dot{\rho}_{R}+4H\rho_{R} & =\Gamma_{\phi}\left(\rho_{\phi}+p_{\phi}\right),\label{eq:rho-R}\\
	\dot{q}_{C}+3Hq_{C} & =\Gamma_{C}\left(q_{C}^{\text{eq}}-q_{C}\right).\label{eq:qc}
\end{align}
Here, $H$ is the Hubble parameter, $V$ is the inflaton potential,
$\Gamma_{\phi}$ is the decay rate of $\phi$ to radiation, $\rho_{\phi}=\frac{1}{2}\dot{\phi}^{2}+V$
and $p_{\phi}=\frac{1}{2}\dot{\phi}^{2}-V$ are the inflaton energy density and pressure, and $\Gamma_{C}$ is the rate of the charge-violating process that drives $q_{C}$ towards its equilibrium value $q_{C}^{\text{eq}}$.
In general, we have $q_{C}^{{\rm eq}} = \chi_{CA}T^{2}\mu_{A}^{{\rm eff}}/6$, with the susceptibilities $\chi_{CA}$ depending on the exact choice of $A$ and $C$ and the temperature scale of chargegenesis. The nonzero entries of $\chi_{CA}$ are, however, typically of $\mathcal{O}(1)$. For definiteness, we will therefore simply work with $q_{C}^{{\rm eq}} = T^{2}\mu_{A}^{{\rm eff}}/6$ in the following.

We parametrize the reaction rate $\Gamma_C$ in a model-independent way in terms of an energy scale $M_C$,
\begin{equation}
	\Gamma_{C}=\frac{T^{2n+1}}{M_{C}^{2n}}\thinspace,\label{eq:Gamma-C}
\end{equation}
where $n$ is a positive integer. Eq.~\eqref{eq:Gamma-C} may stem from
an effective operator ${\cal O}_{C}$ with mass dimension $D=4+n$.  For convenience,
we also define the charge density in a comoving volume, $Q_{C}\equiv q_{C}(a/a_i)^{3}$
and $Q_{C}^{{\rm eq}}\equiv q_{C}^{\text{eq}}(a/a_i)^{3}$ with $a$ the FLRW
scale factor, and $a_i$ denoting the value of $a$ at the first peak of the $\phi$ oscillations. 

Since gravitational chargegenesis is only effective when $\omega$ deviates from $-1$ or $1/3$, the specific form of $V$ in the slow-roll regime is not important in this work, as it only affects inflationary predictions.
While the evolution of $q_{C}$ is insensitive to the slow-roll part of the potential, it does depend on the shape of the potential close to its minimum, which can be modeled by a mass term. For concreteness, we adopt
the following $T$-model potential~\cite{Kallosh:2013hoa,Barman:2022qgt}\footnote{The general form of $T$-model potentials is $V\propto\tanh^{p}(\phi/\sqrt{6})$, which approximates a broad class of models. For instance,
the Starobinsky model approximately corresponds to $p=2$. }
\begin{equation}
	V(\phi)=6\lambda\tanh^{2}\left(\phi/\sqrt{6}\right),\label{eq:-3}
\end{equation}
which  reduces to $V(\phi)\approx\sfrac{1}{2}\,m_{\phi}^{2}\phi^{2}$
with $m_{\phi}=\sqrt{2\lambda}$ at small $\phi$. The slow-roll parameter $\epsilon = \sfrac{1}{2}\,(V'/V)^2$ for this potential is simply $\epsilon=\sfrac{4}{3}\,\text{csch}^{2}(\sqrt{\sfrac{2}{3}}\,\phi)$.
The constant $\lambda$ is determined by $\lambda\approx18\pi^{2}A_{S}/(6N_{e}^{2})$
\cite{Barman:2022qgt} where $N_{e}=55$ is the number of e-folds
and $A_{S}\approx2.1\times10^{-9}$ \cite{Planck:2018jri}. 

\begin{figure}
	\centering
	
	\includegraphics[width=0.49\textwidth]{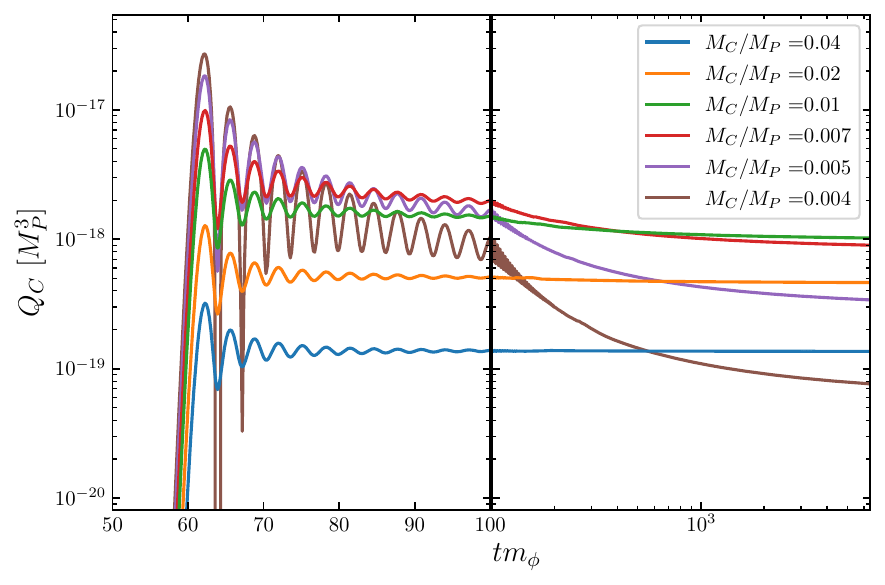}\caption{Numerical solutions of Eq.~\eqref{eq:qc} for different
		values of $M_{C}$. At $tm_{\phi}=100$, we switch to a log scale on the $tm_\phi$ axis. \label{fig:qc}}
\end{figure}

We numerically solve Eqs.~\eqref{eq:phi} to \eqref{eq:qc} starting from a point deep in the slow-roll regime with $\epsilon\ll1$. Specifically, we set the initial point at $\phi=6.13$, corresponding to $N_{e}=55$, continuously evolve the system of equations through the end of slow-roll ($\epsilon=1$), followed by a large number of inflaton oscillations, and stop at a sufficiently large time $t$, when $Q_{C}$ has become constant. In Fig.~\ref{fig:qc},
we present the numerical solutions for $M_{*}=8.22\times10^{-5}$, $\Gamma_{\phi}=1.23\times10^{-7}$, $n=1$, and several values of $M_{C}$. We set $t=0$ at $N_{e}=55$ and denote quantities at the end of slow-roll by a subscript ``$_{\star}$''. For the shown examples, $t_{\star}\approx59.4 \,m_{\phi}^{-1}$, $\phi_{\star}\approx1.21$, and $H_{\star}\approx3.3\times10^{-6}$. 

During the slow-roll phase, $Q_{C}$ cannot be significantly produced because $\omega\approx-1$, which leads to $Q_{C}^{{\rm eq}}\propto\mu_{A}^{{\rm eff}}\approx0$. From the end of slow-roll to $t\sim{\cal O}(100)\,m_{\phi}^{-1}$, $Q_{C}$ is effectively produced, with an oscillatory behavior driven by the oscillations of $\phi$ around the minimum of its potential. The oscillation frequency is of ${\cal O}(m_{\phi})$. During the oscillatory phase, $\phi$ can be effectively viewed as a fluid with $\omega\approx0$ due to $\langle p_{\phi}\rangle/\rho_{\phi}\approx0$, implying an effective matter-dominated phase. As $t$ further increases, $Q_{C}$ asymptotically approaches a constant value. The oscillations in $Q_{C}$ are eventually suppressed for two reasons: 
% \begin{enumerate}[(i)]
\begin{enumerate}[label=(\roman*)]
	\item \label{it:1} The energy density $\rho_{\phi}$ gradually decays to $\rho_{R}$, driving the Universe towards complete radiation domination;

	\item \label{it:2}  The decreasing temperature reduces the rate $\Gamma_{C}$, which hinders $Q_{C}$ from following the rapid oscillations in $Q_{C}^{{\rm eq}}$. Note that for the shown examples, $\Gamma_{C}$ is always well below $m_{\phi}$. 
\end{enumerate}

We now comment on the conditions under which the calculation can be trusted. Some further details can be found in Appendix~\ref{appendix:consistecy}.
The effect of the interaction~\eqref{riccicurrentGeneral} on the cosmological expansion can be neglected if its contribution to the Friedmann equation is negligible, which leads to
\begin{align}
	\label{MstarCondition}
	M_*^2\gg \frac{q_C m_\phi}{M_P^2}.
\end{align}
We verified that this condition is satisfied in our calculations. Furthermore, if the operator~\eqref{riccicurrentGeneral} arises from the low-energy expansion of a fundamental UV-complete theory, then additional constraints must be imposed to ensure that expansion parameters are under control. A necessary condition to suppress higher-dimensional operators arising from a weak-gravitational field expansion is $\mathcal{R}/M_*^2\lesssim 1$. Additionally, $T\sqrt{\mathcal{R}}/M_*^2\lesssim 1$ is required to suppress higher-dimensional derivative operators, see e.g.\ Refs.~\cite{McDonald:2014yfg,McDonald:2015iwt,McDonald:2016ehm,McDonald:2020ghc} and references therein for details. One can readily verify that our choice of parameters respects these constraints both during and after reheating.

\section{Analytical results}
Under certain assumptions, Eqs.~\eqref{eq:phi} to \eqref{eq:qc} can be solved analytically, allowing us to gain some analytical insight into the evolution of relevant quantities. Below we present our analytical results with the detailed derivation relegated to Appendix \ref{sec:derivation}. 

\begin{figure*}
	\centering
	
	\includegraphics[width=0.47\textwidth]{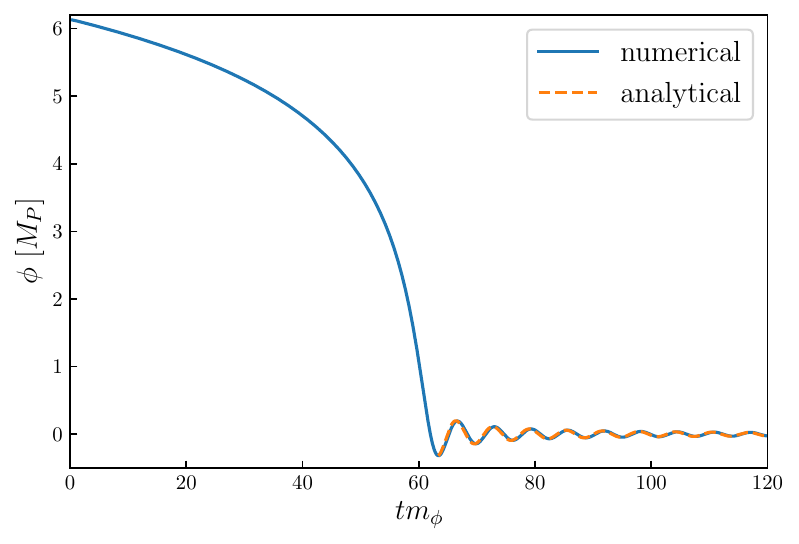}\includegraphics[width=0.49\textwidth]{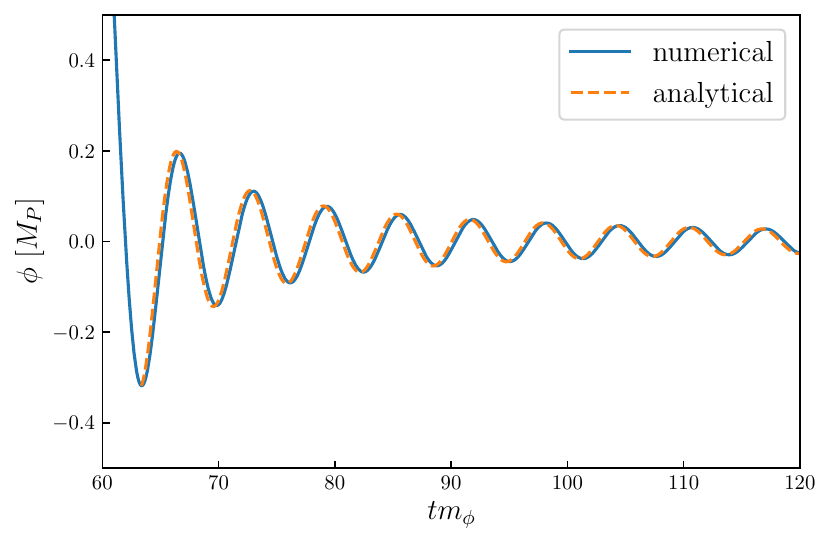}
	
	\caption{The analytical solution in Eq.~\eqref{eq:phi-sol} compared with the
		full numerical solution obtained by solving Eq.~\eqref{eq:phi} with
		the T-model potential in Eq.~\eqref{eq:-3}. The right panel is simply
		a zoom-in view of the left. \label{fig:phi}}
\end{figure*}

First, when $H\gg \Gamma_{\phi}$, we can neglect the effect of  $\Gamma_{\phi}$ on the evolution of $\phi$. 
During this period, we find that the expansion of the Universe can be approximated as follows:
% assuming that the decay rate $\Gamma_{\phi}$ is not too large to affect the evolution of $\phi$ before $H$
\begin{align}
	a & \approx a_{i}\left(1+\frac{3}{2}h_{m}\theta_{t}\right)^{\frac{2}{3}},\label{eq:-24}\\
	H & \approx\frac{2m_{\phi}h_{m}}{3h_{m}\theta_{t}+2}\thinspace.\label{eq:H-approx2}
\end{align}
Here $h_{m}$ and $\theta_{t}$ are two dimensionless quantities:
\begin{equation}
	h_{m}\equiv\frac{H_{i}}{m_{\phi}}\thinspace,\ \ \theta_{t}\equiv m_{\phi}(t-t_{i})\thinspace.\label{eq:hm-def}
\end{equation}
The subscript ${i}$ denotes the moment when $\phi$ reaches the first peak of its oscillations.  When $\phi$ starts oscillating, it approximately follows
\begin{equation}
	\phi\approx\frac{2\phi_{i}}{3h_{m}\theta_{t}+2}\cos\theta_{t}\thinspace,\label{eq:phi-sol}
\end{equation}
which agrees well with the numerical solution. 
Figure~\ref{fig:phi}
shows the accuracy of the above approximate solution for $\phi$ as a function of time for $t\geq t_i$. The numerical
solution (blue curve) in this plot is obtained by solving Eq.~\eqref{eq:phi}
with the T-model potential in Eq.~\eqref{eq:-3} and $\Gamma_{\phi}=0$.

Next, one can compute $\rho_\phi$ straightforwardly from Eq.~\eqref{eq:phi-sol}:
\begin{equation}
	\rho_{\phi}\approx\frac{2m_{\phi}^{2}\phi_{i}^{2}}{\left(3h_{m}\theta_{t}+2\right){}^{2}}\left[1+h_{t}\sin2\theta_{t}+\left(h_{t}\cos\theta_{t}\right)^{2}\right],\label{eq:rho-phi}
\end{equation}
where 
\begin{equation}h_{t}\equiv\frac{3h_{m}}{3h_{m}\theta_{t}+2}\,.
\end{equation}
In Eq.~\eqref{eq:rho-phi},
the three terms are proportional to $h_{t}^{0}$, $h_{t}^{1}$ and
$h_{t}^{2}$ with $h_{t}\ll1$. Hence, $\rho_{\phi}$ does not
exhibit large oscillations as the second and
third oscillatory terms are quickly suppressed at large $t$. However, this is not
the case for $\dot{R}$ computed from $\rho_{\phi}$ and its time
derivatives ($\dot{\rho}_{\phi}$, $\ddot{\rho}_{\phi}$), as we will
show later. Therefore, the oscillatory
terms in Eq.~\eqref{eq:rho-phi} remain important for subsequent calculations. 

It is worth mentioning here that if we treat $\phi$ in the oscillation
phase as a fluid with $\omega_{\phi}=0$, we would obtain an analytical
expression for $\rho_{\phi}$ corresponding to the non-oscillatory
term in Eq.~\eqref{eq:rho-phi}. More specifically, in the fluid approximation,
Eq.~\eqref{eq:phi} can be recast as
\begin{equation}
	\dot{\rho}_{\phi}+\left(3H+\Gamma_{\phi}\right)\left(1+\omega_{\phi}\right)\rho_{\phi}=0\thinspace.\label{eq:rho-phi-eq}
\end{equation}
By setting $\Gamma_{\phi}=\omega_{\phi}=0$ and using Eq.~\eqref{eq:H-approx2}, we obtain
\begin{equation}
	\rho_{\phi}\approx3\left(\frac{2M_{P}m_{\phi}h_{m}}{3h_{m}\theta_{t}+2}\right)^{2},\label{eq:rho-phi-0}
\end{equation}
which is exactly the first term (i.e., the non-oscillating part) in Eq.~\eqref{eq:rho-phi}.

\begin{figure}[t]
	\centering
	
	\includegraphics[width=0.49\textwidth]{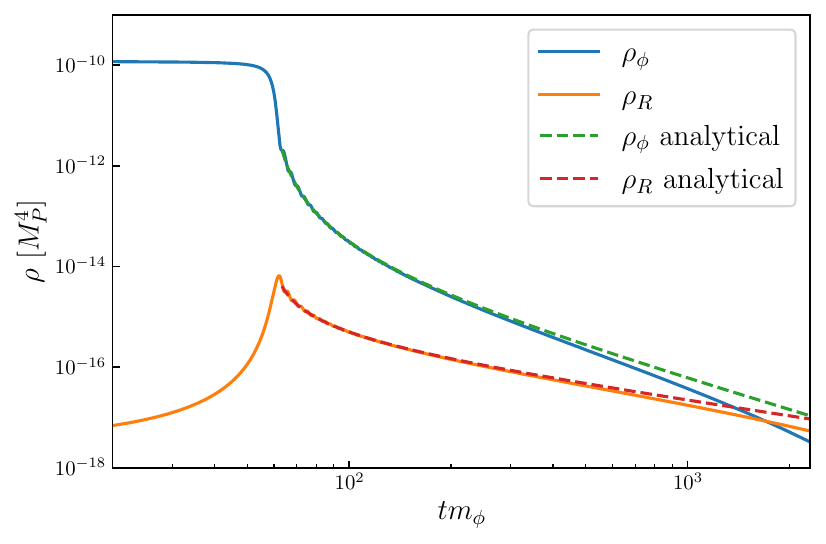}
	
	\caption{Evolution of $\rho_{\phi}$ and $\rho_{R}$. Solid lines are computed
		in the full numerical approach, while dashed lines are obtained using
		Eqs.~\eqref{eq:rho-phi-0} and \eqref{eq:-28}. \label{fig:rho-compare}}
\end{figure}

\begin{figure*}
	\centering
	
	\includegraphics[width=0.48\textwidth]{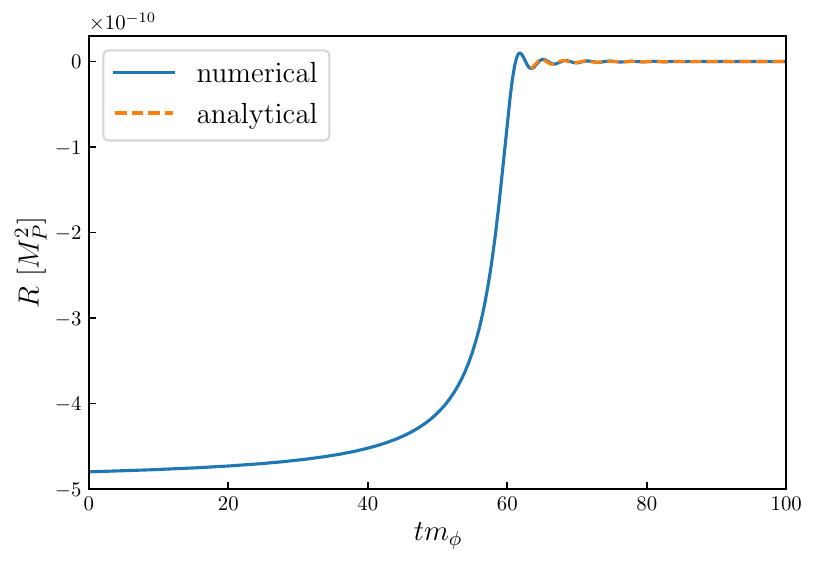}\includegraphics[width=0.49\textwidth]{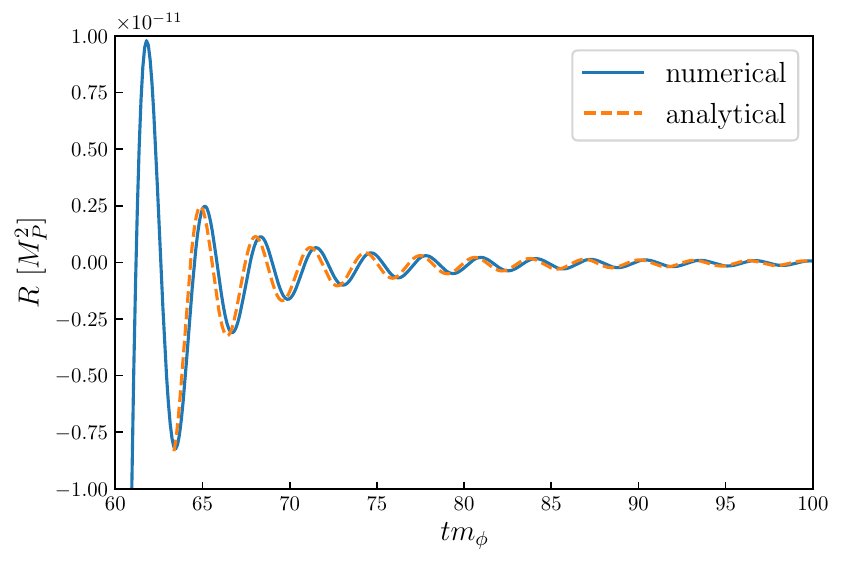}
	
	\includegraphics[width=0.49\textwidth]{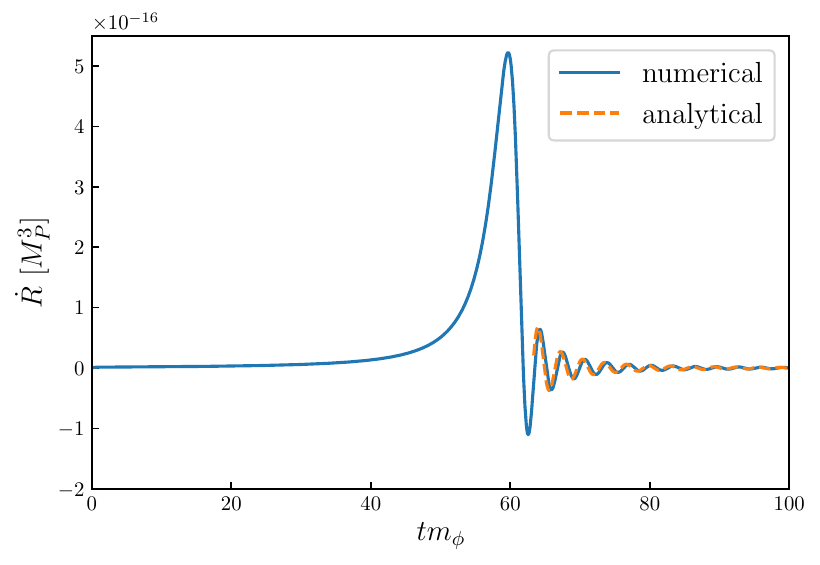}\includegraphics[width=0.49\textwidth]{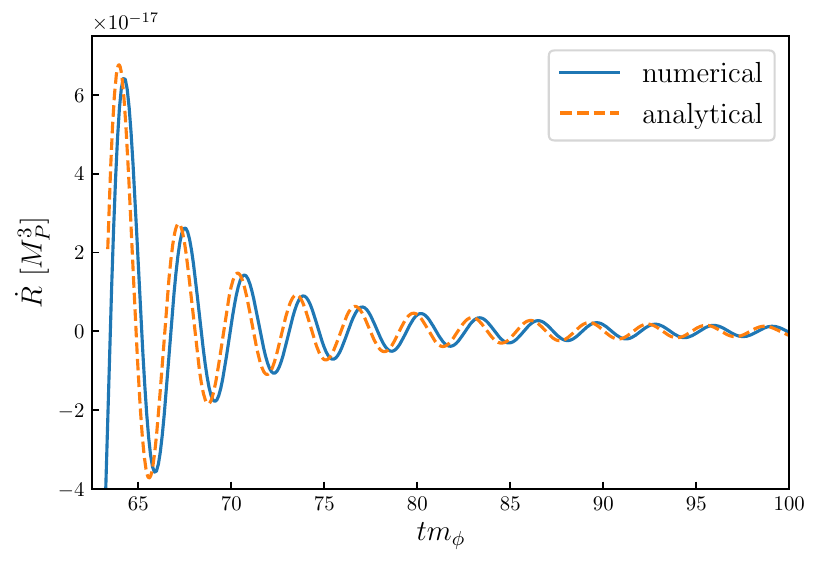}\caption{The analytical expressions in Eqs.~\eqref{eq:-18} and \eqref{eq:-19}
		for $\mathcal{R}$ (upper panels) and $\dot{\mathcal{R}}$ (lower
		panels), compared with the full numerical results obtained by solving
		Eq.~\eqref{eq:phi} with the T-model potential in Eq.~\eqref{eq:-3}.
		The right panels are simply zoom-in views of the left. \label{fig:R-Rdot}}
\end{figure*}

\begin{figure}
	\centering
	
	\includegraphics[width=0.49\textwidth]{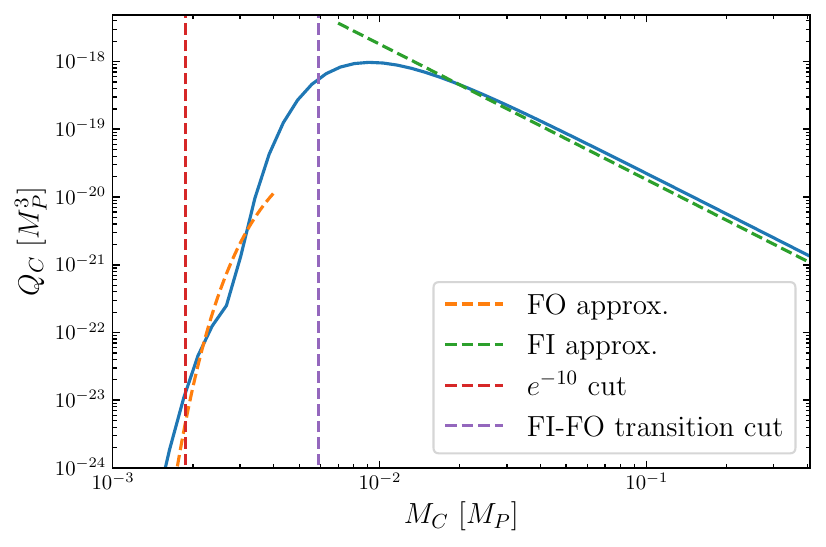}\caption{Final value of $Q_{C}$ as a function of $M_{C}$. The green and
		orange dashed lines represent the FI and FO approximate expressions
		in Eqs.~\eqref{eq:FI} and \eqref{eq:FO}, respectively. The purple
		and red dashed curves indicate the two cuts in Eq.~\eqref{eq:e10-FI-FO}.
		\label{fig:Qc-Mc}}
\end{figure}

From Eq.~\eqref{eq:rho-phi-0}, it is also straightforward to compute $\rho_R$:
\begin{equation}
	\rho_{R}\approx \frac{12\Gamma_{\phi}M_{P}^{2}m_{\phi}h_{m}}{5\left(3h_{m}\theta_{t}+2\right)}\thinspace.\label{eq:-28}
\end{equation}
Note that here we only present the non-oscillating part of $\rho_R$ as the oscillating part is not important to the calculation of $\dot{\mathcal{R}}$. 
% Although $\rho_R$ also contains oscillating parts, they are not important in our subsequent subsequent calculations because 
% $\rho_R$:

In Figure \ref{fig:rho-compare}, we display a comparison of the analytical results in Eqs.~\eqref{eq:rho-phi-0}
and \eqref{eq:-28} with full numerical results, for $\Gamma_{\phi}\approx4.11\times10^{-9}M_{P}$.
As can be seen in the figure, Eqs.~\eqref{eq:rho-phi-0} and
\eqref{eq:-28} describe the evolution of $\rho_{\phi}$
and $\rho_{R}$ accurately before the Universe transitions from $\rho_{\phi}$
domination to $\rho_{R}$ domination.

During the epoch of $\rho_{\phi}$ domination, one can use Eq.~\eqref{eq:rho-phi} to compute $\mathcal{R}$, which reads\footnote{We note here that Eq.~\eqref{eq:-18} is similar to Eq.~(4.22) in \cite{Lozanov:2019jxc},
but our result contains an extra factor ``$+2$'' in the denominator, which yields a more accurate approximation at small $\theta_{t}$.}:
\begin{equation}
	\mathcal{R}\approx-\frac{2m_{\phi}^{2}\phi_{i}^{2}}{M_{P}^{2}\left(3h_{m}\theta_{t}+2\right){}^{2}}\left[1+3\cos(2\theta_{t})\right].\label{eq:-18}
\end{equation}
The time derivative of Eq.~\eqref{eq:-18} gives
\begin{equation}
	\dot{\mathcal{R}}\approx\frac{12m_{\phi}^{3}\phi_{i}^{2}}{M_{P}^{2}\left(3\theta_{t}h_{m}+2\right)^{2}}\left[\sin2\theta_{t}+\frac{h_{m} (1+3\cos2\theta_{t})}{3\theta_{t}h_{m}+2}\right].\label{eq:-19}
\end{equation}

 In Fig.~\ref{fig:R-Rdot},
we plot Eqs.~\eqref{eq:-18} and \eqref{eq:-19}, and compare them
with full numerical results obtained by solving Eq.~\eqref{eq:phi} with
$\Gamma_{\phi}=0$. As is shown in the figure, the analytical expressions
agree well with the numerical results.

% ++++++++++++++
% {\color{red}the previous beginning of sec xxx}

% As is shown in Fig.~\ref{fig:qc}, the production rate and final value of $Q_{C}$ crucially depend on $M_{C}$.

With the analytical result of $\dot{\mathcal{R}}$, we can further study the evolution of $q_C$ analytically. For concreteness, the scope of the following discussion is restricted to charge-violating rates described by~\eqref{eq:Gamma-C} with $n=1$. Generalized expressions for $n\geq 1$ can be found in Appendix~\ref{sec:derivation}. Using a few approximations (see Appendix.~\ref{sec:derivation}), we find that there are two interesting regimes in which the final value of $Q_{C}$ can be given by simple expressions. We refer to them as the freeze-in  (FI) and freeze-out (FO) regimes, to be elucidated below.

{\textbf{Freeze-in regime:}}
The freeze-in regime occurs when the interaction rate $\Gamma_C$ is sufficiently low. 
% {\color{red}The final value of $Q_{C}$ scales like $M_{C}^{-1}$.}
For example, the curves in Fig.~\ref{fig:qc} with $M_{C}\gtrsim0.01\thinspace M_{P}$ are in the freeze-in regime.  In this regime, $Q_{C}$ quickly approaches a constant due to 
% reason (ii)
reason \ref{it:2} discussed previously.
Here, the final value of $Q_{C}$
is determined by the charge produced during the first few inflaton oscillations. We find the following analytical expression  for $Q_{C}$ in this regime,
\begin{equation}
	Q_{C}\approx\frac{m_{\phi}^{13/4}\phi_{i}^{13/4}\Gamma_{\phi}^{5/4}}{M_{D}^{3/4}M_{*}^{2}M_{C}^{2}}\ \ (\text{FI})\thinspace,\label{eq:FI}
\end{equation}
where $M_{D}^{3/4}\approx537M_{P}^{3/4}$ and $\phi_{i}$ denotes
the first peak value of $|\phi|$ during the oscillation phase. 
For $\Gamma_{\phi}\ll H$, we obtain $\phi_{i}\approx0.3M_{P}$, which is approximately a constant since
the slow roll followed by the first oscillation is not significantly affected by $\Gamma_{\phi}$.

{\textbf{Freeze-out regime:}} If $M_{C}^{-1}$ is increased above a certain value, the asymmetry yield during the oscillatory phase will saturate an upper bound expected from the fluid approximation, which is roughly the average $\langle Q_{C}^{{\rm eq}}\rangle$. On the other hand, large $M_{C}^{-1}$ causes $Q_{C}$ to be more tightly coupled to $\langle Q_{C}^{{\rm eq}}\rangle$. When $\langle Q_{C}^{{\rm eq}}\rangle$ is suppressed due to radiation domination ($\omega\approx1/3$), the tight coupling will reduce the final value of $Q_{C}$.  In this regime, we are able to derive 
\begin{equation}
	Q_{C}\approx\frac{\Gamma_{\phi}^{2}m_{\phi}^{2}\phi_{i}^{2}}{M_{F}M_{*}^{2}}\exp\left[-\frac{M_{E}^{3/2}\Gamma_{\phi}^{1/2}}{M_{C}^{2}}\right]\ \ (\text{FO})\thinspace,\label{eq:FO}
\end{equation}
where $M_{E}\approx0.21M_{P}$ and $M_{F}\approx86M_{P}$.

Analytically, one can estimate the value of $M_{C}$ at the transition
between freeze-in and freeze-out and also a lower bound of $M_{C}$
below which $Q_{C}$ is exponentially suppressed---see Appendix~\ref{sec:derivation}. Here we provide two useful cuts:
\begin{equation}
	M_{C}\approx M_{E}^{3/4}\Gamma_{\phi}^{1/4} \times \begin{cases}
		1.0 & \text{for FI--FO cut}\\[2mm]
		0.3 & \text{for }e^{-10}\ \text{cut}
	\end{cases}\thinspace.\label{eq:e10-FI-FO}
\end{equation}
The FI--FO cut indicates the transition while the $e^{-10}$ cut implies
that the exponential in \eqref{eq:FO} leads to a suppression factor
of $e^{-10}$. 

In Fig.~\ref{fig:Qc-Mc}, we show the final value of $Q_{C}$ as
a function of $M_{C}$. One can see that the above analytical estimates
agree well with the numerical result.

% \section{Freeze-in/out and parameter space}
\section{Parameter space}

In gravitational chargegenesis, once the inflation potential and the form of $\Gamma_{C}$ are specified, there are three free (tunable) parameters, namely $\Gamma_{\phi}$, $M_{C}$, and  $M_{*}$. The influence of $M_{*}$ on our results is simple: since the chemical potential in Eq.~\eqref{eq:chem} is proportional to $M_{*}^{-2}$, the final value of $Q_C$ should also be proportional to $M_{*}^{-2}$, as can be verified from Eqs.~\eqref{eq:FI} and \eqref{eq:FO}.
Therefore, we are mainly concerned with the viable parameter space of $\Gamma_{\phi}$ and $M_{C}$ that can lead to successful gravitational chargegenesis. The value of $M_{*}$ is set at $2\times10^{14}\ \text{GeV}$ in the following analysis. This value of $M_*$ respects the constraint in Eq.~\eqref{MstarCondition} and the constraints mentioned in the text below Eq.~\eqref{MstarCondition} in all of the parameter space considered in the following. 

\begin{figure}

\centering
\includegraphics[width=0.49\textwidth]{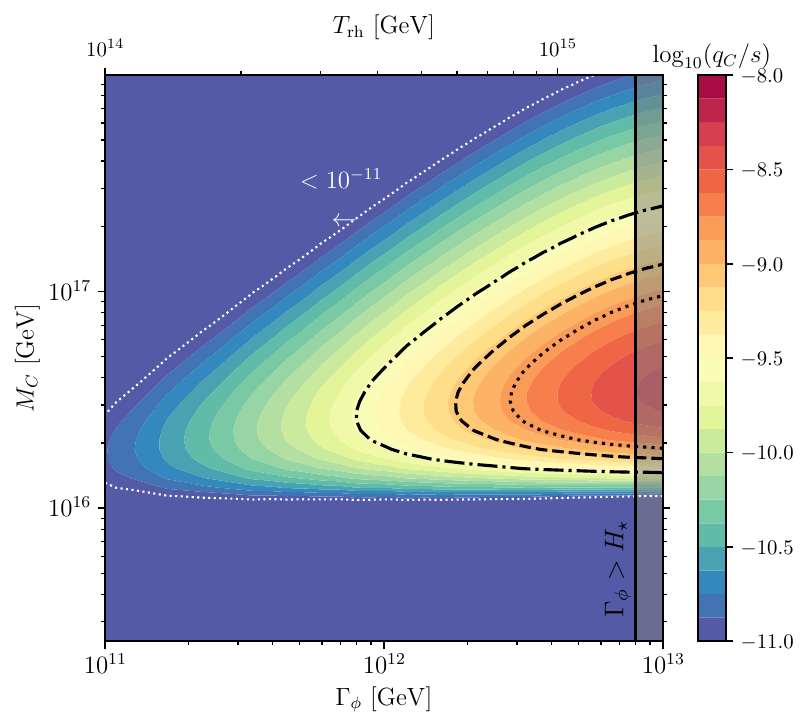}
 
\caption{Viable parameter space of gravitational chargegenesis. The black dashed contour ($q_{C}/s=8.2\times10^{-10}$) can generate the observed baryon asymmetry assuming $|x_{C}|=3/10$, while the dotted and dash-dotted contours assume $|x_{C}|=1/6$ and $1$.
Larger $\Gamma_{\phi}$ generally leads to a higher yield of charge asymmetry, but too large $\Gamma_{\phi}$ would spoil the slow-roll predictions. Hence we cut off the scan when $\Gamma_{\phi}>H_{\star}$, represented by the black solid line and the gray shaded region to the right. \label{fig:contour}}
\end{figure}

In order to identify the viable values of $\Gamma_{\phi}$ and $M_{C}$, we perform a full numerical scan.
% We perform a full numerical 
% scan of the parameter space to identify the range of values of $\Gamma_{\phi}$
% and $M_{C}$ that can lead to successful gravitational chargegenesis.
Fig.~\ref{fig:contour} shows the obtained result in terms of $q_{C}/s$
where $s$ is the entropy density.   The top axis of Fig.~\ref{fig:contour}
indicates the reheating temperature $T_{{\rm rh}}$, which is defined
as the temperature of the Universe at the moment when $\rho_{\phi}=\rho_{R}$.
This value is extracted directly from our numerical solutions while
analytically one expects $T_{{\rm rh}}\propto\Gamma_{\phi}^{1/2}$.
Indeed, in our numerical solutions, we find $T_{{\rm rh}}\approx0.35M_{P}^{1/2}\Gamma_{\phi}^{1/2}$
within the range of $\Gamma_{\phi}$ presented in Fig.~\ref{fig:contour}. 

Fig.~\ref{fig:contour} shows that the yield of charge
asymmetry increases with increasing reheating temperature. However, if $\Gamma_{\phi}$
is too large, $\phi$ would decay even before the slow-roll ends,
alternating the slow-roll paradigm. The Planck 2018 result constrains
the Hubble parameter during inflation to be $H<2.5\times10^{-5}$ at 95\% CL \cite{Planck:2018jri},
implying an upper bound on the reheating temperature, $T_{{\rm rh}}<6.6\times10^{15}$
GeV. If one naively extrapolates $T_{{\rm rh}}\approx0.35M_{P}^{1/2}\Gamma_{\phi}^{1/2}$
to this scale, it can be interpreted as $\Gamma_{\phi}<1.5\times10^{14}$
GeV. However, such a large $\Gamma_{\phi}$ would already exceed $H_{\star}$
and significantly modify the slow-roll evolution. Therefore, we conservatively
set $\Gamma_{\phi}<H_{\star}\approx8\times10^{12}$ GeV as the upper
bound of $\Gamma_{\phi}$ in Fig.~\ref{fig:contour}. 

In wash-in leptogenesis, the baryon asymmetry of the Universe, $Y_{B}\equiv(n_{B}-n_{\overline{B}})/s$,
is related to  $q_{C}$ by 
\begin{equation}
	Y_{B}=\frac{28}{79}x_{C}\frac{q_{C}}{s}\thinspace,\label{eq:YB}
\end{equation}
where $x_{C}$ is the coefficient relating the $q_{C}$ asymmetry
to the $B-L$ asymmetry. For electron asymmetry ($q_{e}$) produced
before the electron Yukawa interaction reaches equilibrium around $T\sim 10^{5}$
GeV~\cite{Bodeker:2019ajh}, we have $x_{C}=-3/10$~\cite{Domcke:2020quw}. Hence to generate
the observed value $Y_{B}\approx8.7\times10^{-11}$ (see e.g.~\cite{Davidson:2008bu}),
we need $|q_{C}|/s\approx8.2\times10^{-10}$. This corresponds to
the black dashed line in Fig.~\ref{fig:contour}. For other charge asymmetries, we refer to Tab.~II of Ref.~\cite{Domcke:2020quw}, where
$|x_{C}|$ typically varies from $1/6$ to $1$. The other two black
contours demonstrate such variations. 

We have verified that by setting $n>1$ in~\eqref{eq:Gamma-C}, a charge asymmetry sufficient to explain the BAU can be produced with lower values of $M_C$ than in the $n=1$ scenario. However, the allowed range for $M_C$ becomes narrower compared to those restricted by the black curves in Fig.~\ref{fig:contour}. Setting $n>1$ also opens up the possibility of producing a sufficient charge asymmetry at lower reheating temperatures. Finally, we have also verified that for $T$-model potentials that can be approximated as $\sim \phi^p$, $p=6,\,8,\,10,...$ during reheating, one can potentially reproduce the BAU in gravitational chargegenesis scenarios with significantly lower reheating temperatures than those indicated in Fig.~\ref{fig:contour}. However, a consistent study of reheating potentials beyond the quadratic requires accounting for fragmentation of the inflaton condensate (see e.g.~\cite{Lozanov:2016hid,Lozanov:2017hjm}) and is left for future work. 

\section{Summary and conclusions}
In this paper, we proposed a framework for chargegenesis, which, via wash-in leptogenesis, can explain the baryon asymmetry of the Universe. The two main ingredients are a gravitational interaction that dynamically generates effective chemical potentials and a charge-violating interaction that drives the plasma to a state with nonzero charge asymmetry $q_C$. The charge associated with the gravitational interaction and the charge being violated do not have to be the same. In contrast to previous gravitational baryogenesis frameworks, the present framework allows for the generated charge to not even violate $B-L$ provided one right-handed neutrino is still active at a time when $q_C$ is non vanishing. 
  %In particular, we extended the scope of previous gravitational baryogenesis ideas by pointing out that the charge associated with the gravitational interaction and the charge that is being violated do not have to be the same. In the present framework, the generated charge does not even have to violate $B-L$ provided as long as one right-handed neutrino is still active at a time when $q_C$ is non vanishing. 

As an application of the framework, we considered a concrete realization of gravitational chargegenesis after a period of slow-roll inflation where the inflaton oscillates around the minimum of its potential. We focused on the simplest and most generic setup where the inflaton oscillates in a quadratic potential and provided a detailed time-resolved description of how $q_C$ evolves throughout reheating and beyond. We showed that a sufficient charge to explain the BAU can be generated while staying consistent with experimental constraints. As this mechanism, at least its most minimal realizations as considered here, generally requires a high inflation scale it also implies a large primordial tensor-to-scalar ratio in the cosmic microwave background power spectrum that could be detected by upcoming experiments. This renders gravitational chargegenesis an exciting possibility to pursue theoretically and experimentally. \\

%%%%%%%%%%%%%%%%%%%%%%%%%%%%%%%%%%%%%%%%%%%%%%%%%%%%%%%%%%%%%%%%%%%%%%%%%%%%%%%%%%%%%%%%%%%%%%%%%%%

\subsection*{Acknowledgements}

%%%%%%%%%%%%%%%%%%%%%%%%%%%%%%%%%%%%%%%%%%%%%%%%%%%%%%%%%%%%%%%%%%%%%%%%%%%%%%%%%%%%%%%%%%%%%%%%%%%

M.\,A.\,M.\ thanks Hooman Davoudiasl for a pleasant conversation about gravitational baryogenesis, and Yong Xu for discussions about reheating. The work of K.\,S.\ is supported by Deutsche Forschungsgemeinschaft (DFG) through the Research Training Group (Graduiertenkolleg) 2149: Strong and Weak Interactions -- from Hadrons to Dark Matter. M.\,A.\,M.\ acknowledges support from the DFG Collaborative Research Centre ``Neutrinos and Dark Matter in Astro- and Particle Physics'' (SFB 1258). The work of X.\,J.\,X is supported in part by the National Natural Science Foundation of China under grant No.~12141501 and also by the CAS Project for Young Scientists in Basic Research (YSBR-099). 

%%%%%%%%%%%%%%%%%%%%%%%%%%%%%%%%%%%%%%%%%%%%%%%%%%%%%%%%%%%%%%%%%%%%%%%%%%%%%%%%%%%%%%%%%%%%%%%%%%%

\small

\bibliographystyle{utphys}
%\bibliography{arxiv_1}
\bibliography{references}

%%%%%%%%%%%%%%%%%%%%%%%%%%%%%%%%%%%%%%%%%%%%%%%%%%%%%%%%%%%%%%%%%%%%%%%%%%%%%%%%%%%%%%%%%%%%%%%%%%%%

\newpage
\onecolumngrid
\newpage

%%%%%%%%%%%%%%%%%%%%%%%%%%%%%%%%%%%%%%%%%%%%%%%%%%%%%%%%%%%%%%%%%%%%%%%%%%%%%%%%%%%%%%%%%%%%%%%%%%%%

% \renewcommand{\thesection}{S\arabic{section}}
% \renewcommand{\theequation}{S\arabic{equation}}
% \renewcommand{\thefigure}{S\arabic{figure}}
% \renewcommand{\thetable}{S\arabic{table}}
% \setcounter{equation}{0}
% \setcounter{figure}{0}
% \setcounter{table}{0}
% \setcounter{page}{1}
% \setcounter{section}{0}

%%%%%%%%%%%%%%%%%%%%%%%%%%%%%%%%%%%%%%%%%%%%%%%%%%%%%%%%%%%%%%%%%%%%%%%%%%%%%%%%%%%%%%%%%%%%%%%%%%%%

% \begin{center}
% \textbf{\large Supplemental Material: Gravitational chargegenesis}\\
% Martin A Mojahed, Kai Schmitz, Xun-Jie Xu
% \end{center}

%%%%%%%%%%%%%%%%%%%%%%%%%%%%%%%%%%%%%%%%%%%%%%%%%%%%%%%%%%%%%%%%%%%%%%%%%%%%%%%%%%%%%%%%%%%%%%%%%%%%
\appendix

\section{Expansion of the Universe}
\label{appendix:consistecy}
In this appendix, we provide some details about the consistency of our framework. The theory under consideration has the following schematic action, 
\begin{align}
    \label{Sint}
    S=\int\,d^4x\,\sqrt{-g}\left[\frac{{\cal R}}{2}M_P^2\left(1-\frac{2\,\partial_\mu J_C^\mu}{M_*^2M_P^2}\right)+\mathcal{L}_M\right]+...
\end{align}
where we have discarded a total derivative and additional gravitational operators that may appear in a more complete model description. Here $\mathcal{L}_{M}$ denotes the non-gravitational interactions in the Lagrangian. 

The standard Einstein equation is recovered by discarding the current operator and varying~\eqref{Sint} with respect to the metric. The Friedmann equation, which determines the expansion of the Universe, is simply the $(0,0)$ component of Einstein's equation. In the presence of the current operator, the Friedmann equation is modified. A simple calculation reveals that the modification is negligible if
\begin{align}
    \frac{\dot{q}_C}{M_*^2M_P^2}\ll 1.
\end{align}
The charge $q_C$ is negligible during inflation, and starts to rapidly oscillate with a period set by the inflaton mass $m_\phi$ during the early stages of reheating, as shown in Fig.~\ref{fig:qc}. Hence, requiring the current-operator under study to have negligible impact on the expansion of the Universe leads to the constraint in~\eqref{MstarCondition}.

\section{Derivation of the analytical results}
\label{sec:derivation}

In this appendix, we present the explicit derivation
of the analytical results used in this work. 

\subsection{Analytical expressions for \boldmath{$\phi$}, \boldmath{$\rho_{\phi}$}, and \boldmath{$\rho_{R}$}}

The Universe is in a matter-dominated era when $\phi$ oscillates at the bottom of its potential. During the matter-dominated era, the Hubble parameter is approximately given by $H\approx\frac{2}{3t}$.
Since $\frac{2}{3t}$ is divergent at $t\to0$, we regulate it by shifting the beginning of matter domination away from $t=0$
to $t=t_{i}$:
\begin{equation}
	H\approx\frac{1}{H_{i}^{-1}+\frac{3(t-t_{i})}{2}}\thinspace,\label{eq:H-approx1}
\end{equation}
such that the transition from an inflationary epoch
to matter domination happens around $t=t_{i}$ and $H_{i}$ denotes
the Hubble parameter at this transition. In practice, we assume that this transition happens when $\phi$ reaches its first peak, and define the initial moment ``$i$'' at exactly the peak.  
% We denote the corresponding values of $\phi$, $H$, and $a$ at
% this moment by $\phi_{i}$, $H_{i}$, and $a_{i}$, respectively.
By this definition, we have $\dot{\phi}(t_{i})=0$, which implies $\rho_{\phi}(t_{i})=m_{\phi}^{2}\phi_{i}^{2}/2$
and $H_{i}\approx M_{P}^{-1}\sqrt{\rho_{\phi}/3}\approx M_{P}^{-1}m_{\phi}|\phi_{i}|/\sqrt{6}$,
assuming $\rho_{R}\ll\rho_{\phi}$ at $t=t_{i}$. 
Therefore, $h_{m}$ and
$\phi_{i}$, defined in Eq.~\eqref{eq:hm-def},  are related by 
\begin{equation}
	h_{m}=\frac{|\phi_{i}|}{\sqrt{6}M_{P}}\thinspace,\ \ \text{or}\ \ |\phi_{i}|=\sqrt{6}h_{m}M_{P}.\label{eq:-20}
\end{equation}
From Eq.~\eqref{eq:H-approx1}, it is straightforward to compute the
scale factor by solving $\dot{a}/a=H$. Writing the approximate expressions
of $H$ and $a$ in  terms of $h_{m}$ and $\theta_{t}$, we obtain the results in Eqs.~\eqref{eq:-24} and \eqref{eq:H-approx2}.
% \begin{align}
% 	a & \approx a_{i}\left(1+\frac{3}{2}h_{m}\theta_{t}\right)^{\frac{2}{3}},\label{eq:-24}\\
% 	H & \approx\frac{2m_{\phi}h_{m}}{3h_{m}\theta_{t}+2}\thinspace.\label{eq:H-approx2}
% \end{align}

Substituting Eq.~\eqref{eq:H-approx2} into Eq.~\eqref{eq:phi} and
using the quadratic expansion of $V$, one can solve Eq.~\eqref{eq:phi}
analytically in the limit of $\Gamma_{\phi}\to0$. The result is given in Eq.~\eqref{eq:phi-sol}.
% \begin{equation}
% 	\phi\approx\frac{2\phi_{i}}{3h_{m}\theta_{t}+2}\cos\theta_{t}\thinspace,\label{eq:phi-sol}
% \end{equation}
% which agrees well with the numerical solution. 
% Figure \ref{fig:phi}
% shows the accuracy of the above approximate solution for $\phi$ as a function of time for $t\geq t_i$. The numerical
% solution (blue curve) in this plot is obtained by solving Eq.~\eqref{eq:phi}
% with the T-model potential in Eq.~\eqref{eq:-3} and $\Gamma_{\phi}=0$.
Then from Eq.~\eqref{eq:phi-sol}, one can derive an analytical approximation for $\rho_\phi$, as given in Eq.~\eqref{eq:rho-phi}.

Next, we consider Eq.~\eqref{eq:rho-R} for $\rho_{R}$. If $\dot{\rho}_{R}\ensuremath{\ll4H\rho_{R}}$,
we can ignore the $\dot{\rho}_{R}$ term in Eq.~\eqref{eq:rho-R}
and obtain 
\begin{equation}
	\rho_{R}\approx\frac{\Gamma_{\phi}\left(1+\omega_{\phi}\right)\rho_{\phi}}{4H}\approx\frac{3m_{\phi}\Gamma_{\phi}h_{m}M_{P}^{2}}{2\left(3h_{m}\theta_{t}+2\right)}\thinspace,\ \ (\text{assuming }\text{\ensuremath{\dot{\rho}_{R}\,}\ensuremath{\ensuremath{\ll}4H\ensuremath{\rho_{R}}}})\thinspace,\label{eq:-8}
\end{equation}
where we used Eq.~\eqref{eq:rho-phi-0} and $\omega_{\phi}=0$.
However, we find that $\dot{\rho}_{R}\ensuremath{\ll4H\rho_{R}}$
is actually not a very good approximation as Eq.~\eqref{eq:-8} would imply
\begin{equation}
	\dot{\rho}_{R}\approx-\frac{3h_{m}m_{\phi}}{3h_{m}\theta_{t}+2}\rho_{R}\thinspace,\ \ (\text{assuming }\text{\ensuremath{\dot{\rho}_{R}\,}\ensuremath{\ensuremath{\ll}4H\ensuremath{\rho_{R}}}}).\label{eq:-29}
\end{equation}
For $\theta_{t}=0$ and $h_{m}\theta_{t}\gg1$, Eq.~\eqref{eq:-29}
gives $|\dot{\rho}_{R}|\approx3H_{i}\rho_{R}$ and $|\dot{\rho}_{R}|\approx\rho_{R}/t\approx3H\rho_{R}/2$,
respectively. Both are smaller than $4H\rho_{R}$ but not much smaller
than $4H\rho_{R}$. To improve the approximation, we substitute
Eq.~\eqref{eq:-29} into Eq.~\eqref{eq:rho-R} to account for
the contribution from the $\dot{\rho}_{R}$ term. The resulting equation reads
\begin{equation}
	\left(-\frac{3h_{m}m_{\phi}}{3h_{m}\theta_{t}+2}+4H\right)\rho_{R}=\Gamma_{\phi}\left(1+\omega_{\phi}\right)\rho_{\phi}\thinspace,\label{eq:-30}
\end{equation}
which upon substitution of $\omega_{\phi}=0$ yields Eq.~\eqref{eq:-28}.

\subsection{Analytical expressions for \boldmath{$\mathcal{R}$} and \boldmath{$\dot{\mathcal{R}}$}}

In the $(+,---)$ metric convention, the Ricci scalar $\mathcal{R}$ in the FRW Universe is given
by
\begin{align}
	\mathcal{R} & =-6\left[\frac{\ddot{a}}{a}+\left(\frac{\dot{a}}{a}\right)^{2}\right]=-6\left[\dot{H}+2H^{2}\right].\label{eq:-10}
\end{align}
The time-derivative of the Ricci scalar is readily obtained by applying the second Friedmann equation,
\begin{equation}
	\dot{\mathcal{R}}=\frac{3\rho\dot{\omega}}{M_{P}^{2}}+\frac{\sqrt{3}(1+\omega)(1-3\omega)}{M_{P}^{3}}\rho^{3/2}\thinspace.\label{eq:-11}
\end{equation}
By substituting Eq.~\eqref{eq:H-approx2} into Eq.~\eqref{eq:-10} and evaluating the time derivative,
we obtain the following analytical results 
\begin{align}
	\mathcal{R} & \approx-\frac{12h_{m}^{2}m_{\phi}^{2}}{(3h_{m}\theta_{t}+2)^{2}}\thinspace,\label{eq:-13}\\
	\dot{\mathcal{R}} & \approx\frac{72h_{m}^{3}m_{\phi}^{3}}{(3h_{m}\theta_{t}+2)^{3}}\thinspace,\label{eq:-14}
\end{align}
which are valid during reheating.
Eqs.~\eqref{eq:-13} and \eqref{eq:-14} should be considered as leading-order approximations that do not account for oscillations. To arrive at a more accurate description, we now incorporate the oscillatory terms in $\rho_{\phi}$
into $\mathcal{R}$ and $\dot{\mathcal{R}}$, as shown in the following. 

Given the explicit expression of $\rho$ and its derivatives ($\dot{\rho}$,
$\ddot{\rho}$), Eq.~\eqref{eq:-10} can be used to directly compute
$\mathcal{R}$. This is most easily seen by recalling that $H=M_{P}^{-1}\sqrt{\rho/3}$ and $\dot{H}=M_{P}^{-1}d(\sqrt{\rho/3})dt$,
i.e.~all relevant quantities can be written in terms of $\rho$ and its time derivatives.
Hence, starting from
\begin{align}
	\mathcal{R} & =-\frac{4\rho+\dot{\rho}/H}{M_{P}^{2}}\thinspace,\label{eq:-17}
\end{align}
and taking $\rho\approx\rho_{\phi}$, substituting Eqs.~\eqref{eq:rho-phi}
and \eqref{eq:H-approx2} into Eq.~\eqref{eq:-17}, neglecting a few
subdominant terms such as the $h_{t}^{2}$ term in Eq.~\eqref{eq:rho-phi}, we eventually obtain Eqs.~\eqref{eq:-18} and \eqref{eq:-19}. If we average out the $\sin2\theta_{t}$ and $\cos2\theta_{t}$ terms
in Eq.~\eqref{eq:-19}, $\dot{\mathcal{R}}$ reduces exactly to Eq.~\eqref{eq:-14}, as expected.

\subsection{Freeze-in and freeze-out values of \boldmath{$Q_{C}$}}
Next, we consider analytical approximations of $Q_C$ for the freeze-in and the freeze-out regimes.
By defining $Q_{C}\equiv q_{C}a^{3}/a_i^{3}$ and $Q_{C}^{{\rm eq}}\equiv q_{C}^{{\rm eq}}a^{3}/a_i^{3}$,
one can rewrite Eq.~\eqref{eq:qc} as
\begin{equation}
	\frac{dQ_{C}}{dt}=\Gamma_{C}Q_{C}^{\text{eq}}-\Gamma_{C}Q_{C}\thinspace.\label{eq:-22-1}
\end{equation}
According to Appendix B in \cite{Schmitz:2023pfy}, the solution for
this equation can be formally written as
\begin{equation}
	Q_{C}(t)=G(t)\int_{0}^{t}\frac{\Gamma_{C}(t')Q_{C}^{\text{eq}}(t')}{G(t')}dt'\thinspace,\ \ G(t)\equiv\exp\left[-\int_{0}^{t}\Gamma_{C}\left(t'\right)dt'\right].\label{eq:-36}
\end{equation}
In the limit of $\Gamma_{C}\to0$, i.e. the freeze-in regime, we have
$G\to1$ and thus
\begin{equation}
	Q_{C}\approx\int\Gamma_{C}Q_{C}^{\text{eq}}dt\approx\int\dot{R}Kdt\thinspace,\ \ K\equiv\frac{(a/a_i)^3}{6M_{*}^{2}}\frac{T^{2n+3}}{M_{C}^{2n}}\thinspace.\label{eq:-23}
\end{equation}
Here, the temperature $T$ can be determined from $\rho_{R}$,
\begin{equation}
	T=\left(\frac{30}{\pi^{2}g_{\star}}\rho_{R}\right)^{1/4}=c_{g}\rho_{R}^{1/4}\thinspace\ \ \text{with}\ \ c_{g}\equiv\left(\frac{30}{\pi^{2}g_{\star}}\right)^{1/4}\approx0.41\thinspace,\label{eq:T-rhoR}
\end{equation}
where the number of effective relativistic degrees of freedom $g_{\star}$ in the standard model is $g_\star=106.75$ at the temperature scales relevant to our work. By combining the analytical results \eqref{eq:-24}, \eqref{eq:-28}, 
and \eqref{eq:-19},  we get 
\begin{align}
    \dot{R}K&\approx \frac{ m_\phi^3\phi_i^2}{2M_*^2M_C^{2n}M_P^2}\left(\frac{h_m}{3h_m\theta_t+2}\right)\left(\frac{72\,\Gamma_\phi M_P^2m_\phi h_m}{\pi^2g_\star(3h_m\theta_t+2)}\right)^{\frac{2n+3}{4}}\left[1+3\cos2\theta_{t}+\frac{3\theta_{t}h_{m}+2}{h_{m}}\sin2\theta_{t}\right].\label{eq:-31}
\end{align}
By neglecting oscillatory terms and integrating from $t=t_{i}$ (corresponding to $\theta_{t}=0$) to $t=\infty$,
we obtain 
\begin{align}
 \int_{\slashed{s}\slashed{c}}\dot{R}Kdt&\approx \frac{2}{3(2n+3)}\frac{ m_\phi^2\phi_i^2}{M_*^2M_C^{2n}M_P^2}\left(\frac{36\,\Gamma_\phi M_P^2m_\phi h_m}{\pi^2g_\star}\right)^{\frac{2n+3}{4}},\label{eq:-33}
\end{align}
where ``$\slashed{s}\slashed{c}$'' reminds us that we have neglected
the $\sin2\theta_{t}$ and $\cos2\theta_{t}$ terms in Eq.~\eqref{eq:-31}.
To quantify the deviation of this simplified result from the full result, which accounts for oscillations in $\dot{R}K$, we define the following quantity
\begin{align}
    r_{h}&\equiv\frac{\int_{\slashed{s}\slashed{c}}\dot{R}Kdt}{\int\dot{R}Kdt}=\frac{\int_{0}^{\infty}d\theta_{t}\left(3h_{m}\theta_{t}+2\right)^{-(2n+7)/4}}{\int_{0}^{\infty}d\theta_{t}\left(3h_{m}\theta_{t}+2\right)^{-(2n+7)/4}\left[1+3\cos2\theta_{t}+\frac{3\theta_{t}h_{m}+2}{h_{m}}\sin2\theta_{t}\right]}\thinspace,\label{eq:-34}
\end{align}
which is only a function of $h_{m}$.  For $h_{m}\ll1$, we find
that $r_{h}$ varies within a quite narrow range, typically  between
$0.34$ and $0.35$ for $n=1$. Finally, by combining the expressions above, we obtain the freeze-in value 
of $Q_{C}$: 
\begin{align}
    Q_{C}&\approx\frac{1}{r_{h}}\frac{2}{3(2n+3)}\frac{ m_\phi^2\phi_i^2}{M_*^2M_C^{2n}M_P^2}\left(\frac{36\,\Gamma_\phi M_P^2m_\phi h_m}{\pi^2g_\star}\right)^{\frac{2n+3}{4}}\thinspace\ \ \ (\text{for  freeze-in})\thinspace.\label{eq:-32}
\end{align}

When $\Gamma_{C}$ is too large, the approximations used to derive
the freeze-in value are no longer valid but one can analytically estimate
$Q_{C}$ in the freeze-out regime. In this regime, we assume that $Q_{C}$ is tightly coupled
to $Q_{C}^{{\rm eq}}$ until the Universe enters the $\rho_{R}$-dominated
era, in which $Q_{C}^{{\rm eq}}$ quickly vanishes because $\rho_{\phi}$
decays exponentially in the subsequent evolution. Under this assumption,
we are only concerned with the evolution of $Q_{C}$ starting from
the point when $\rho_{R}=\rho_{\phi}$. 

By equating Eq.~\eqref{eq:rho-phi-0} to \eqref{eq:-28}, we obtain an approximate
value for $\theta_{t}$ at the time when $\rho_{R}$ becomes equal to $\rho_{\phi}$,
\begin{equation}
	\theta_{t}\approx\frac{5m_{\phi}h_{m}-2\Gamma_{\phi}}{3\Gamma_{\phi}h_{m}}\thinspace\ \ (\text{for }\rho_{R}=\rho_{\phi})\thinspace.\label{eq:theta-cross}
\end{equation}
Using Eq.~\eqref{eq:theta-cross} and assuming that the oscillating terms
in $Q_{C}^{{\rm eq}}$ are negligible, we obtain 
\begin{align}
	\left\langle Q_{C}^{{\rm eq}}\right\rangle &\approx\frac{\sqrt{3} c_{g}^{2}\Gamma_{\phi}^{2}m_{\phi}^{2}\phi_{i}^{2}}{25M_{P}M_{*}^{2}}\thinspace\ \ (\text{at }\rho_{R}=\rho_{\phi})\thinspace,\label{eq:Qc-x}
\end{align}
where $c_{g}$ has been defined in Eq.~\eqref{eq:T-rhoR}. Once $\rho_{R}$
increases across $\rho_{\phi}$, $Q_{C}^{{\rm eq}}$ will drop exponentially
as $t$ becomes comparable to $\Gamma_{\phi}^{-1}$. Meanwhile, $Q_{C}$
decreases at a much lower rate determined by $\Gamma_{C}$. The subsequent part
of the evolution is no longer sensitive to $Q_{C}^{{\rm eq}}$, which
can safely be set to zero. Eventually $Q_{C}$ will reach a stable value
when $\Gamma_{C}\ll H$. Under this assumption, we apply the formalism
in Eq.~\eqref{eq:-36} to the period starting from $\rho_{R}=\rho_{\phi}$
and integrate $t$ out to infinity. This results in 

\begin{align}
    Q_{C}&\approx\left\langle Q_{C}^{{\rm eq}}\right\rangle \exp\left[-\frac{\sqrt{3}}{2n-1}\left(\frac{2\sqrt{3}}{5}\right)^{\frac{2n-1}{2}}\frac{c_{g}^{2n+1}M_{P}^{\frac{2n+1}{2}}\Gamma_{\phi}^{\frac{2n-1}{2}}}{M_{C}^{2n}}\right]\thinspace\ \ \ (\text{for freeze-out})\thinspace.\label{eq:FO-a}
\end{align}
If $M_{C}$ is very small, then $Q_{C}$ in Eq.~\eqref{eq:FO-a} would
be exponentially suppressed. To estimate the magnitude of $M_{C}$
for when the exponential suppression becomes highly efficient, we solve the following equation,
\begin{equation}
	\frac{Q_{C}}{\left\langle Q_{C}^{{\rm eq}}\right\rangle }=e^{-10}\thinspace,\label{eq:-38}
\end{equation}
which gives 
\begin{align}
    M_{C}&=\left(\frac{\sqrt{3}}{10(2n-1)}\right)^{\frac{1}{2n}}\left(\frac{2\sqrt{3}}{5}\right)^{\frac{2n-1}{4n}}\sqrt{c_g^2M_P\Gamma_\phi}\left(\frac{c_g^2M_P}{\Gamma_\phi}\right)^{\frac{1}{4n}}.\label{eq:Mc-e-10}
\end{align}
We refer to this $M_{C}$ value as the $e^{-10}$ cut, or $M_C^{\rm{cut}}$, below which
$Q_{C}$ is expected to be suppressed by at least a factor of $e^{-10}\sim10^{-5}$
compared to $Q_{C}^{\text{eq}}$ at $\rho_{R}=\rho_{\phi}$. 

One can estimate the transition from the freeze-in to the freeze-out regime by introducing a transition value $M_C^{\rm{trans}}$ for which the argument of the exponential in \eqref{eq:FO-a} equals $-1$. This definition implies the following relation
\begin{align}
    \label{eq:trans-cut}
    M_C^{\rm{trans}}=10^{\frac{1}{2n}}M_C^{\rm{cut}},
\end{align}
which manifests the fact that the FO window becomes narrower for increasing $n$, as mentioned in the main text.

\section{Slow-roll parameters}

When adopting the T-model potential in Eq.~\eqref{eq:-3}, the slow-roll
parameters can be obtained in compact analytical forms,
\begin{align}
	\epsilon & \equiv\frac{M_{P}^{2}}{2}\left(\frac{V'}{V}\right)^{2}=\frac{4}{3}\text{csch}^{2}\left(\sqrt{\frac{2}{3}}\frac{\phi}{M_{P}}\right),\label{eq:sr-1}\\
    \eta & \equiv M_{P}^{2}\left(\frac{V''}{V}\right)=\frac{4}{3}\left[2-\cosh{\left( \sqrt{\frac{2}{3}} \frac{\phi}{M_{P}} \right) }\right]\text{csch}^2\left(\sqrt{\frac{2}{3}}  \frac{\phi}{M_{P}} \right).\label{eq:sr-2}
\end{align}
The number of e-folds is given by 
\begin{equation}
	N_{e}\equiv\frac{1}{\sqrt{2}M_{P}}\int_{\phi_e}^{\phi}\frac{d\phi'}{\sqrt{\epsilon(\phi')}}=\frac{3}{4}\cosh\left(\sqrt{\frac{2}{3}}\frac{\phi}{M_P}\right)-\frac{3}{4}\cosh\left(\sqrt{\frac{2}{3}}\frac{\phi_e}{M_P}\right),\label{eq:sr-3}
\end{equation}
where $\phi_{e}$ denotes $\phi$ at the end of inflation.

%%%%%%%%%%%%%%%%%%%%%%%%%%%%%%%%%%%%%%%%%%%%%%%%%%%%%%%%%%%%%%%%%%%%%%%%%%%%%%%%%%%%%%%%%%%%%%%%%%% 

\end{document}